\documentclass[a4paper]{article}

\usepackage{INTERSPEECH2020}
\usepackage{multirow}
\usepackage{makecell}
\usepackage[flushleft]{threeparttable}
\usepackage{hyperref}
\usepackage{amsmath}
\usepackage{soul,color}

\title{MatchboxNet: 1D Time-Channel Separable Convolutional Neural Network Architecture for Speech Commands Recognition}
\name{Somshubra Majumdar, Boris Ginsburg\thanks{Preprint. Submitted to INTERSPEECH.}}

\address{NVIDIA, Santa Clara, USA }
\email{\{smajumdar,bginsburg\}@nvidia.com} 

\begin{document}

\maketitle

\begin{abstract}
We present an \textit{MatchboxNet} - an end-to-end neural network for speech command recognition. MatchboxNet is a deep residual network composed from blocks of 1D time-channel separable convolution, batch-normalization, ReLU and dropout layers.  MatchboxNet reaches state-of-the art accuracy on the Google Speech Commands dataset while having significantly fewer parameters than similar  models. The small footprint of MatchboxNet makes it an attractive  candidate for devices with limited computational resources. The model is highly scalable, so model accuracy can be improved with modest additional memory and compute. Finally, we show how intensive data augmentation using an auxiliary noise dataset improves robustness in the presence of background noise.

\end{abstract}
\noindent\textbf{Index Terms}: key word spotting, speech commands recognition, deep neural networks, depth-wise separable convolution

\section{Introduction}

We present MatchboxNet, a new compact, end-to-end neural network for keyword spotting (KWS) specifically designed for devices with low computational and memory resources.
MatchboxNet builds on the QuartzNet architecture \cite{kriman2019quartznet}. It consists of a stack of blocks with residual connections \cite{he2015}. Each block is composed from 1D time-channel separable convolutions (these are similar to 2D depth-wise separable convolutions \cite{chollet2017xception,kaiser2017depthwise}), batch normalization, ReLU and dropout layers. 

This paper makes the following contributions:
\begin{enumerate}
  \item An end-to-end neural model for speech command recognition based on 1D time-channel separable convolutions
  \item The model achieves state-of-the-art accuracy on Google Speech command datasets \cite{warden2018speech} but requires significantly fewer parameters than models which achieve similar accuracy. 
  \item The model scales well with the number of parameters. 
  \item A methodology to improve the model's robustness to background speech and noise.
\end{enumerate}

\section{Related Work}
Neural network (NN)-based systems for Automatic Speech Recognition (ASR) have a long history, spearheaded by  Time Delay Neural Networks (TDNN) for isolated word recognition \cite{Waibel1989, lang1990}. TDNN and Recurrent NNs (RNNs) were first used together with Hidden Markov Models (HMMs) in hybrid systems, where NN was used only for phonetic classification \cite{Bengio1992, Robinson1994, Hermansky2000}. 

Rapid progress in deep learning for ASR \cite{Graves2004, Graves2005, Hinton2012} triggered research in end-to-end NN-based models for  KWS. In 2015 Sainath and Parada proposed a convolutional NN for a small-footprint KWS \cite{Sainath2015convolution}. Their model was composed of two convolutional layers, max-pooling in the temporal dimension, linear, and soft-max layers.
Following the success of ResNets \cite{he2015} in computer vision, Qian et al. \cite{qian2016very} applied ResNets for ASR. Arik et al. \cite{arik2017convolutional} suggested Convolutional-RNN, which combined the strengths of convolutional layers and recurrent layers to exploit long-range context.

The introduction of the Google Speech Command dataset \cite{warden2018speech} in 2018 accelerated research in KWS and resulted in variety of new NN-based models, including deep residual networks (\cite{Tang2018},  \cite{choi2019temporal}), special RNN with weight sharing \cite{kusupatiFastGRNN2018}, an RNN-Transducer with attention \cite{he2019streaming}, and CNN with dilated convolutions and gating mechanisms \cite{coucke2019dilatedgating}.

\section{MatchboxNet Architecture}

\begin{figure}[t]
  \centering
  \includegraphics[width=\linewidth]{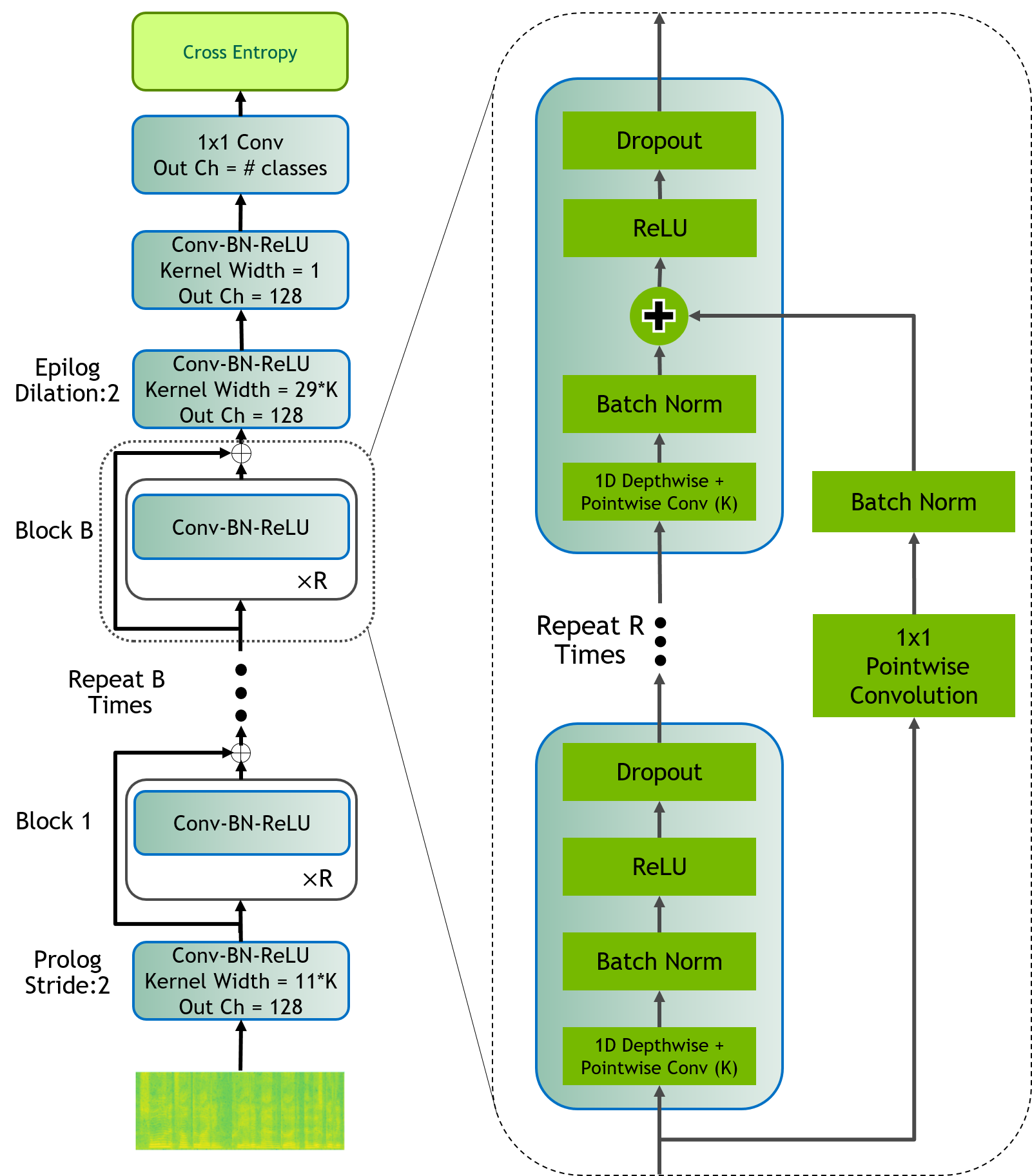}
  \caption{MatchboxNet $B$x$R$x$C$ model: $B$ - number of blocks, \quad $R$ - number of sub-blocks, $C$ - the number of channels.}
  \label{fig:quartznet_arch}
\end{figure}

The MatchboxNet architecture is based on the QuartzNet end-to-end  convolutional NN for ASR~\cite{kriman2019quartznet}. Similar to QuartzNet, MatchboxNet uses 1D time-channel separable convolutions to reduce model size versus regular 1D convolutions. 

A MatchboxNet-$B$x$R$x$C$ model has $B$ residual blocks. Each block has $R$ sub-blocks. All sub-blocks in a block have the same number of output channels $C$ (see Fig.~\ref{fig:quartznet_arch}).
A basic sub-block consists of a 1D-time-channel separable convolution, 1x1 pointwise convolutions, batch norm, ReLU, and dropout. The 1D-time-channel separable convolution has $C$ filters with a kernel of the size $k$.
All models have four additional sub-blocks: one prologue layer -- `Conv1' before the first block, and three epilogue sub-blocks  (`Conv2', `Conv3', and `Conv4') before the final soft-max layer - see Figure~\ref{fig:quartznet_arch}) for details.

For example, the complete architecture for MatchboxNet-3x2x64 (B=3 blocks, R=2 sub-block per block, C=64 channels) is shown in the Table~\ref{tab:QuartzNetParams}:

{\renewcommand{\arraystretch}{1.1}
\begin{table}[!h]
\caption{MatchboxNet-3x2x64 model has B=3 blocks, each black has R=2 time-channel separable convolutional sub-blocks with C=64 channels, plus 4 additional sub-blocks: prologue - Conv1, and epilogue -  Conv2, Conv3, Conv4).}
\label{tab:QuartzNetParams}
\centering
\scalebox{0.8}{
\begin{tabular}{c c c c c} 
 \toprule
  \textbf{Block} &
  \textbf{\# Blocks} &
  \textbf{\thead{\# Sub\\Blocks}}  &
  \textbf{\thead{\# Output\\Channels}} &
  \textbf{Kernel} \\
 \midrule
 Conv1&1 & 1 & 128 & 11 \\
 B1   &1 & 2 & 64 & 13 \\
 B2   &1 & 2 & 64 & 15 \\
 B3   &1 & 2 & 64 & 17 \\
 Conv2 & 1 & 1 & 128 & 29, \textit{dilation=2}\\
 Conv3 & 1 & 1 & 128 & 1 \\
 Conv4 &1 &  1 & \# classes & 1 \\
 Soft-max &\\
 Cross-entropy &\\
 \bottomrule
\end{tabular}
}
\end{table}
}

\section{Experiments}
We train MatchboxNet on the Google Speech Commands Dataset \cite{warden2018speech}. The dataset has two versions which we denote by v1 and v2. Version 1 has 65,000 utterances from various speakers, each utterance is 1 second long. Each of these utterances belongs to one of 30 classes corresponding to common words like “Yes”, “No”, "Go", "Stop", "Left", "Down", numerical digits, etc. 
Version 2 has 105,000 utterances, each 1 second long, belonging to one of 35 classes. 
We re-balanced both training datasets so all classes will have the same number of samples by duplication of random samples.\footnote{One can use cross-entropy loss with class based weighing instead of re-balancing.}

\subsection{Training Methodology}

First, the input audio wave is converted into sequence of 64  mel-frequency cepstral coefficients (MFCC) calculated from 25ms windows with a 10ms overlap. We perform symmetric padding of the temporal dimension with zeros to fixed length of 128 feature vectors per sample. 

Next, the input is augmented with time shift perturbations in the range of $T=[-5, 5]$ milliseconds and white noise with magnitude $[-90, -46]$ dB. In addition, we applied SpecAugment \cite{park2019} with 2 continuous time mask of size $[0, 25]$ time steps, and 2 continuous frequency mask of size $[0, 15]$ frequency bands. We also used SpecCutout \cite{devries2017specutout}, with 5 rectangular masks with time and frequency dimensions similar to used in SpecAugment. 

All models are trained with the NovoGrad optimizer \cite{novograd2019}, with $\beta_1 = 0.95$ and $\beta_2 = 0.5$. We utilize the Warmup-Hold-Decay learning rate schedule as in \cite{he2019bag} with a warm-up ratio of 5\%, a hold ratio of 45\%, and a polynomial (2nd order) decay for the remaining 50\% of the schedule.  We use a maximum learning rate of 0.05 and a minimum learning rate of 0.001. We also incorporate weight decay of 0.001. 
We train all models for 200 epochs using mixed precision  \cite{micikevicius2017mixed} on 2 V-100 GPUs with a batch size of 128 per GPU. 
All experiments were carried out using the NeMo toolkit \cite{nemo2019} and plan to make all code necessary to reproduce these results available.

\subsection{Results}

Comparing with other published results, MatchboxNet-3x1x64 and MatchboxNet-3x2x64 obtain state-of-the-art (SOTA) accuracy on the Google Speech Commands dataset v1 and close to the SOTA on dataset v2, while requiring significantly fewer parameters than other models 
(see Table~\ref{tab:QuartzNet_Scores_v1} and Table~\ref{tab:QuartzNet_Scores_v2}).
For comparison we used the following models:
\begin{itemize}
    \item DenseNet-BC: a variant of ResNets with dense connectivity in between layers of each block \cite{huang2016}. An intermediate point-wise convolution layer applied prior to the convolution block acts as a "bottleneck (B)" layer to reduce number of parameters. The number of channels in the convolutional layer can be reduced via a "compression (C)" factor.
    \item EdgeSpeechNet: ResNet-like deep residual ConvNet  optimized for edge devices \cite{lin2018edgespeechnets}.
    \item Harmonic Tensor 2D-CNN: triangular band-pass filters of the \textit{n}-th harmonic of center frequencies, are extracted and concatenated into a Harmonic Tensor of dimensionality $H \times F \times T$ (harmonic $\times$ frequency $\times$ time) which is then passed into a simple 2D-Convolutional NN \cite{won2020harmonic}.
    \item `Embedding + Head': the acoustic embedding model with multiple heads is pre-trained to distinguish between various keyword groups on 200 million 2-second audio clips from YouTube. These heads are discarded after pre-training, and a single head is used to fine-tune the embedding model on the downstream task \cite{lin2020training}.
\end{itemize}

{\renewcommand{\arraystretch}{1.1}
\begin{table}[!h]
\caption{MatchboxNet on Google Speech Commands dataset v1, the accuracy is averaged over 5 trials (95\% Confidence Interval).}
\label{tab:QuartzNet_Scores_v1}
\centering
\scalebox{0.8}{
\begin{tabular}{c c c c} 
 \toprule
 \textbf {Model} & \textbf{\thead{\# Parameters, K}} & \textbf{Accuracy, \%} & \textbf{Reference} \\
 \midrule
 ResNet-15 & 238 & 95.8 $\pm$ 0.351 & \cite{Tang2018} \\
 DenseNet-BC-100 & 800 & 96.77 & \cite{li2020feature} \\
 EdgeSpeechNet-A & 107 & 96.80 & \cite{lin2018edgespeechnets} \\
 \midrule
 MatchboxNet-3x1x64 & 77 & 97.21 $\pm$ 0.067 & \\
 MatchboxNet-3x2x64 & 93 & 97.48 $\pm$ 0.107 & \\
 \bottomrule
\end{tabular}
}
\end{table}
}

{\renewcommand{\arraystretch}{1.1}
\begin{table}[!h]
\caption{MatchboxNet on Google Speech Commands dataset v2, the accuracy is averaged over 5 trials (95\% Confidence Interval).}
\label{tab:QuartzNet_Scores_v2}
\centering
\scalebox{0.8}{
\begin{tabular}{c c c c} 
 \toprule
 \textbf {Model} & \textbf{\thead{\# Parameters, K}} & \textbf{Accuracy, \%} & \textbf{Reference} \\
 \midrule
  Attention RNN          & 202 & 94.30 & \cite{de2018neural}   \\
  Harmonic Tensor 2D-CNN & -    & 96.39 & \cite{won2020harmonic} \\
  "Embedding + Head" Model & 385   & 97.7 & \cite{lin2020training} \\
 \midrule
  MatchboxNet-3x1x64 & 77 & 96.91 $\pm$ 0.101 &  \\
  MatchboxNet-3x2x64 & 93 & 97.21 $\pm$ 0.072 &  \\
  MatchboxNet-6x2x64 & 140 & 97.37 $\pm$ 0.110 &  \\
 \bottomrule
\end{tabular}
}
\end{table}
}

\subsection{Model Scaling}
We study the model scalability on the Google Speech Commands dataset v2 using MatchboxNet-3x2x64 as baseline. We scale model up using two methods: increase the depth $B\times R$ or increase the number of channels $C$. We found that both methods work in a similar way -- the accuracy increases with model size until we hit $\approx97.6\%$ (Table.~\ref{tab:MatchboxNet_scaling}).\footnote{We analyzed the remaining misclassified samples, and found that most of them are very hard to recognize, even for humans.}

{\renewcommand{\arraystretch}{1.1}
\begin{table}[!h]
\caption{Scaling up MatchboxNet depth and number of channels, Speech Commands Dataset v2}
\label{tab:MatchboxNet_scaling}
\centering
\scalebox{0.8}{
\begin{tabular}{c c c c c } 
 \toprule
 \textbf{B} & \textbf{R}& \textbf{C} & \textbf{\thead{\# Parameters, K}} &
 \textbf{Accuracy,\%} \\
 \midrule
 3 & 2 & 64 & 93 &  97.21 \\
 \midrule
 3 & 3 & 64 & 109 & 97.36 \\
 3 & 4 & 64 & 125 & 97.17 \\
 3 & 5 & 64 & 149 & 97.37 \\
 \midrule
 4 & 2 & 64 &  109 & 97.20 \\
 5 & 2 & 64 &  124 & 97.31 \\
 6 & 2 & 64 &  140 & 97.55 \\
 \midrule
 3 & 2 & 80  & 118 & 97.44 \\
 3 & 2 & 96  & 145 & 97.41 \\
 3 & 2 & 112 & 177 & 97.63 \\
 \bottomrule
\end{tabular}
}
\end{table}
}

\section{Model Robustness to Noise}
To improve the robustness of MatchboxNet in the presence of noise, we retrained the model with background noise designed to interfere with speech signal. We construct a background noise dataset using audio samples from the \textit{Freesound} database \cite{font2013freesound}.  We partition each of these audio samples into segments of 1 second each, with no overlap between segments. Following this methodology, we obtain close to 55,000 noise samples.

\subsection{Training with Noise Augmentation}
We train MatchboxNet-3x1x64 by augmenting all training samples with randomly sampled noise segments. We scale the signal to noise ratio (SNR) randomly between 0 to 50 dB. In cases where the noise segment has a shorter duration than the training sample, we randomly augment a sub-segment of the training sample. The model accuracy on clean data is similar to the baseline model trained with basic augmentation only (Table.~\ref{tab:Noise_augmented_training}).
{\renewcommand{\arraystretch}{1.1}
\begin{table}[!h]
\caption{MatchboxNet-3x1x64 trained with additional background speech and noise augmentation, Google Speech Commands dataset v2. Accuracy (\%) is averaged over 5 trials (95\% confidence interval).}
\label{tab:Noise_augmented_training}
\centering
\scalebox{0.8}{
\begin{tabular}{c c c c} 
 \toprule
 \textbf {Model} & \textbf{Augmentation} &  \textbf{Accuracy, \%} \\
 \midrule
 MatchboxNet 3x1x64 & basic  & 96.91 $\pm$ 0.101 \\
 MatchboxNet 3x1x64 &  + background speech and noise  & 97.05 $\pm$ 0.099 \\
 \bottomrule
\end{tabular}
}
\end{table}
}

\begin{figure}[ht!]
  \centering
  \includegraphics[width=0.85\linewidth]{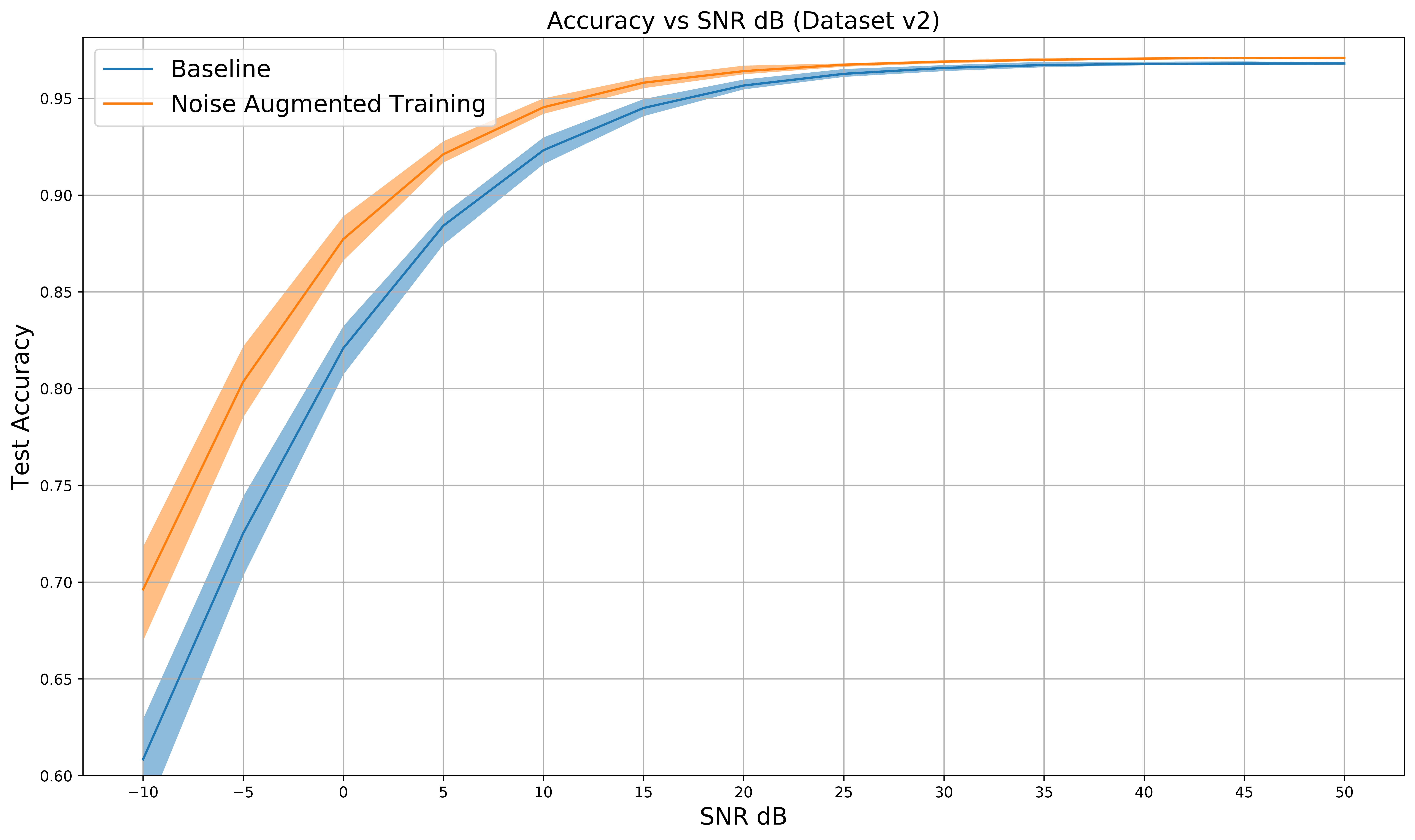}
  \caption{MatchboxNet-$3\times1\times64$ trained with background noise augmentation, Speech Commands dataset v2. Accuracy vs SNR.}
  \label{fig:noise_augmentation_eval_v2}
\end{figure}

In order to evaluate the model robustness to environmental noise and background speech, we test the model with different noise conditions with SNR from -10 dB to +50 dB. We evaluate each test sample with 10 different randomly sampled noise segments, and compute the average accuracy over the entire test set. The model trained with additional noise augmentation is significantly more robust to external noise, even when the noise signal is much higher in amplitude than the noise used during training
(Fig.~\ref{fig:noise_augmentation_eval_v2}).

\begin{figure}[ht!]
  \centering
  \includegraphics[width=0.85\linewidth]{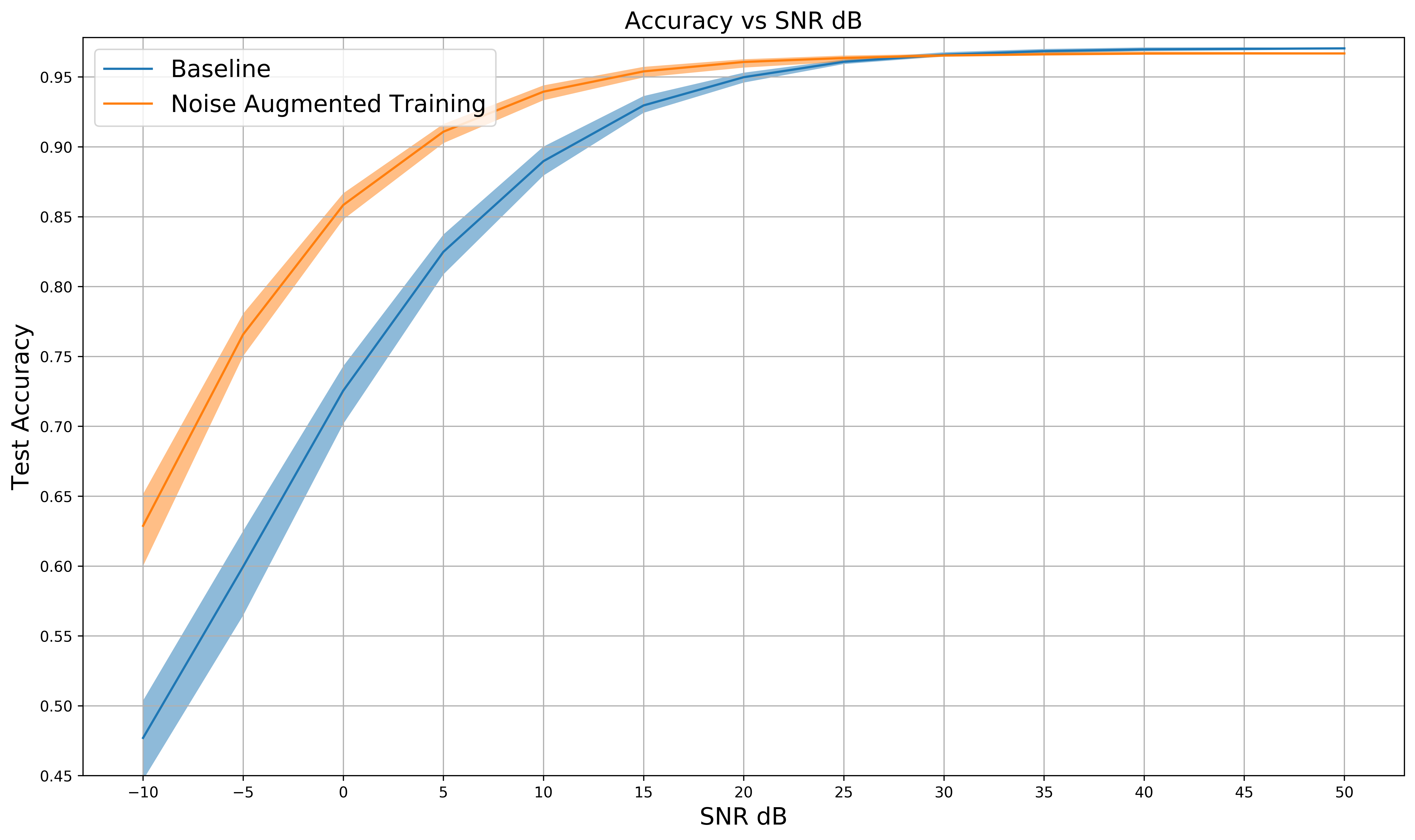}
  \caption{MatchboxNet-$3\times1\times64$ trained with additional background speech and noise augmentation, expanded Google Speech Commands dataset v2. Accuracy vs SNR.}
  \label{fig:noise_detection_eval}
\end{figure}

\subsection{Speech Commands Recognition with Background Speech and Noise Detection}

To use a keyword spotting model in a continuous audio stream, it should be able to differentiate speech commands from the background speech or noise. For this,  we added roughly 3500 samples for environmental noise  and similar number of  background speech samples from Freesound database to the training set. We re-trained a MatchboxNet-3x1x64 model to classify all  original commands plus two additional classes - `background noise' and `background voice'. The model accuracy on the expanded speech commands datasets is shown in Table~\ref{tab:Noise_detection_training}. Training with additional background speech and noise augmentation significantly improves the model robustness to noise (Fig.~\ref{fig:noise_detection_eval}).

{\renewcommand{\arraystretch}{1.1}
\begin{table}[ht!]
\caption{MatchboxNet-$3\times1\times64$ trained with additional background speech and noise augmentation, expanded Speech Commands dataset. Accuracy (\%) is averaged over 5 trials (95\% confidence interval). }
\label{tab:Noise_detection_training}
\centering
\scalebox{0.8}{
\begin{tabular}{c c c c} 
 \toprule
 \textbf {Model} & \textbf{Dataset} & \textbf{\thead{\# Parameters}} & \textbf{Accuracy, \%} \\
 \midrule
 MatchboxNet-3x1x64 & v1 & 77K & 96.88 $\pm$ 0.073 \\
 MatchboxNet-3x1x64 & v2 & 77K & 96.97 $\pm$ 0.071 \\
 \bottomrule
\end{tabular}
}
\end{table}
}

\subsection{Robustness To Noise With Model Scaling}

\begin{figure}[htb!]
  \centering
  \includegraphics[width=0.85\linewidth]{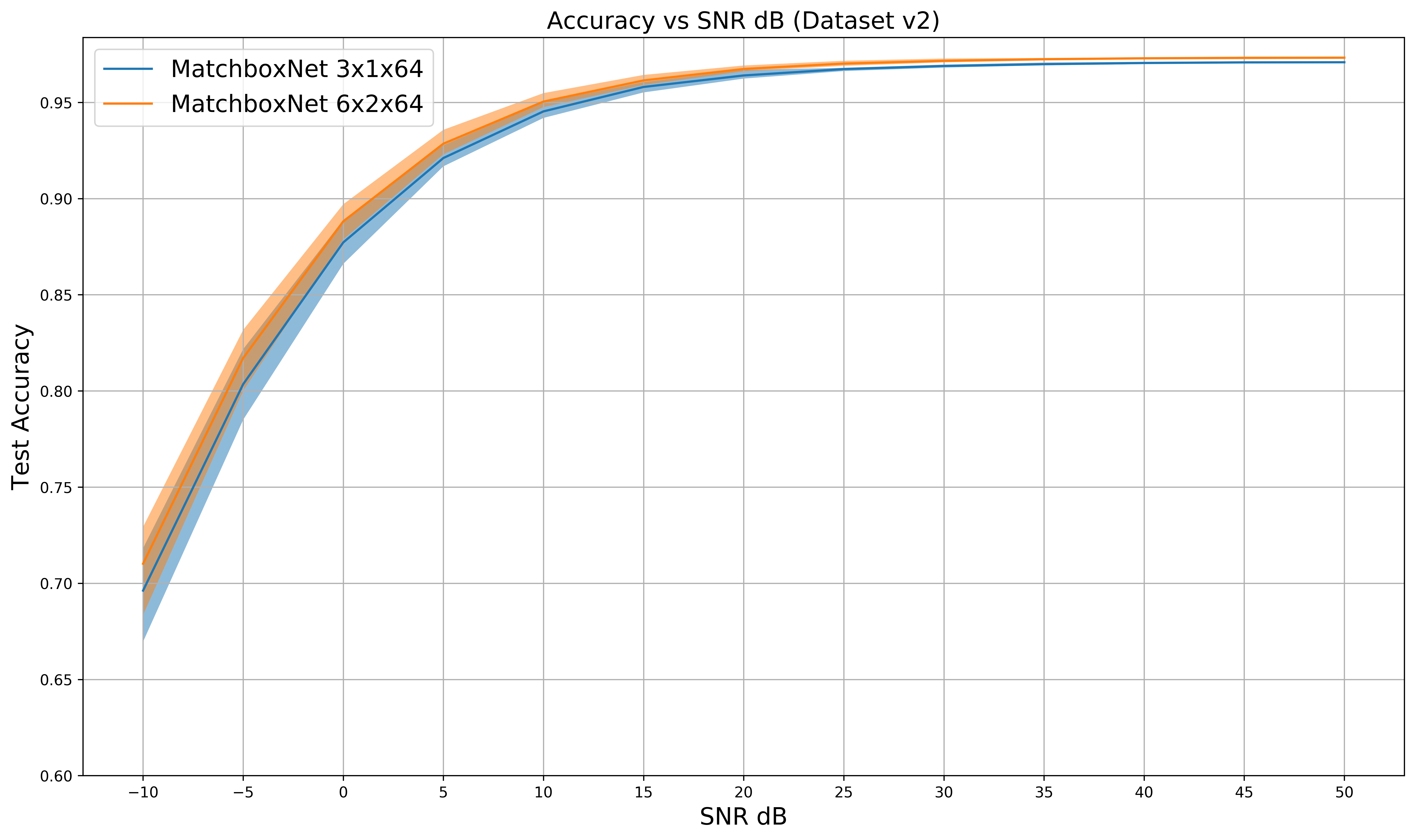}
  \caption{MatchboxNet-$3\times1\times64$ and MatchboxNet-$6\times2\times64$ trained with additional background speech and noise augmentation. Accuracy vs SNR.}
  \label{fig:noise_scaling_fig}
\end{figure}

We further evaluate the relative robustness of larger MatchboxNet models to environmental noise and background speech. We train two models, MatchboxNet-3x1x64 and 6x2x64 with the exact same noise augmentation scheme as described above. We then evaluate the models on the unseen test set, perturbed by 10 random noise samples per test sample and compute the average accuracy. While both models are highly robust to external noise, MatchboxNet-6x2x64 consistently outperforms the smaller MatchboxNet-3x1x64 (see Table~\ref{tab:Noise_Scaling} and Figure~\ref{fig:noise_scaling_fig})

{\renewcommand{\arraystretch}{1.1}
\begin{table}[!h]
\caption{MatchboxNet-$3\times1\times64$ and MatchboxNet-$6\times2\times64$ trained with additional background speech and noise augmentation. Accuracy (\%) is averaged over 10 trials with random noise.}
\label{tab:Noise_Scaling}
\centering
\scalebox{0.8}{
\begin{tabular}{c c c c c c c c c} 
 \toprule
 \textbf {Model} & \multicolumn{7}{c}{\textbf{SNR (in dB)}} \\
 & \textbf{-10} & \textbf{0} & \textbf{10} & \textbf{20} & \textbf{30} & \textbf{40} & \textbf{50} \\
 \midrule
 3x1x64 & 69.62 & 87.21 & 94.53 & 96.40 & 96.89 & 97.05 & 97.09 \\
 6x2x64 & 71.02 & 88.81 & 95.04 & 96.74 & 97.16 & 97.29 & 97.33 \\
 \bottomrule
\end{tabular}
}
\end{table}
}

\section{Conclusions}
In this paper, we present MatchboxNet, a new end-to-end deep neural network architecture for efficient recognition of speech commands on devices with limited computational and memory resources. MatchboxNet is a deep residual network composed from 1D time-channel separable convolution, batch-norm layers, ReLU and dropout layers.
The model has state-of-the-art accuracy on the Google Speech Commands v1 dataset with significantly fewer parameters than models with similar accuracy. MatchboxNet is  scalable, allowing it to be deployed on devices with different memory and compute capabilities. By using intensive data augmentation with auxiliary background noise during training, we have shown the model can be made very robust with respect to background noise.

\section{Acknowledgments}
We would like to thank NVIDIA AI Applications team for the help and valuable feedback.

\bibliography{Jasperbib}

\begin{thebibliography}{10}
\def\url#1{}
\csname url@samestyle\endcsname
\providecommand{\newblock}{\relax}
\providecommand{\bibinfo}[2]{#2}
\providecommand{\BIBentrySTDinterwordspacing}{\spaceskip=0pt\relax}
\providecommand{\BIBentryALTinterwordstretchfactor}{4}
\providecommand{\BIBentryALTinterwordspacing}{\spaceskip=\fontdimen2\font plus
\BIBentryALTinterwordstretchfactor\fontdimen3\font minus
  \fontdimen4\font\relax}
\providecommand{\BIBforeignlanguage}[2]{{%
\expandafter\ifx\csname l@#1\endcsname\relax
\typeout{** WARNING: IEEEtran.bst: No hyphenation pattern has been}%
\typeout{** loaded for the language `#1'. Using the pattern for}%
\typeout{** the default language instead.}%
\else
\language=\csname l@#1\endcsname
\fi
#2}}
\providecommand{\BIBdecl}{\relax}
\BIBdecl

\bibitem{kriman2019quartznet}
S.~Kriman \emph{et~al.}, ``{QuartzNet}: deep automatic speech recognition with
  {1D} time-channel separable convolutions,'' \emph{arXiv:1910.10261}, 2019.

\bibitem{he2015}
K.~He, X.~Zhang, S.~Ren, and J.~Sun, ``Deep residual learning for image
  recognition,'' \emph{arXiv:1512.03385}, 2015.

\bibitem{chollet2017xception}
F.~Chollet, ``Xception: Deep learning with depthwise separable convolutions,''
  in \emph{CVPR}, 2017, pp. 1251--1258.

\bibitem{kaiser2017depthwise}
L.~Kaiser, A.~Gomez, and F.~Chollet, ``Depthwise separable convolutions for
  neural machine translation,'' \emph{arXiv:1706.03059}, 2017.

\bibitem{warden2018speech}
P.~Warden, ``Speech commands: A dataset for limited-vocabulary speech
  recognition,'' \emph{arXiv:1804.03209}, 2018.

\bibitem{Waibel1989}
A.~Waibel, T.~Hanazawa, G.~Hinton, K.~Shirano, and K.~Lang, ``A time-delay
  neural network architecture for isolated word recognition,'' \emph{ICASSP},
  1989.

\bibitem{lang1990}
K.~Lang, A.~Waibel, and G.~Hinton, ``A time-delay neural network architecture
  for isolated word recognition,'' \emph{Neural Networks}, 1990.

\bibitem{Bengio1992}
Y.~Bengio, R.~De~Mori, G.~Flammia, and R.~Kompe, ``Global optimization of a
  neural network-hidden {Markov} model hybrid,'' \emph{IEEE Transactions on
  Neural Networks, 3(2), 252–259}, 1992.

\bibitem{Robinson1994}
T.~Robinson, M.~Hochberg, and S.~Renals, ``{IPA}: improved phone modelling with
  recurrent neural networks,'' \emph{ICASSP}, 1994.

\bibitem{Hermansky2000}
H.~Hermansky, D.~Ellis, and S.~Sharma, ``Tandem connectionist feature
  extraction for conventional hmm systems,'' \emph{ICASSP}, 2000.

\bibitem{Graves2004}
A.~Graves, D.~Eck, N.~Beringer, and J.~Schmidhuber, ``Biologically plausible
  speech recognition with {LSTM} neural nets,'' in \emph{Biologically Inspired
  Approaches to Advanced Information Technology. BioADIT}, 2004.

\bibitem{Graves2005}
A.~Graves and J.~Schmidhuber, ``Framewise phoneme classification with
  bidirectional {LSTM} and other neural network architectures,'' \emph{Neural
  Networks, vol. 18}, pp. 602–--610, 2005.

\bibitem{Hinton2012}
G.~Hinton \emph{et~al.}, ``Deep neural networks for acoustic modeling in speech
  recognition,'' \emph{IEEE Signal Processing Magazine}, 2012.

\bibitem{Sainath2015convolution}
T.~Sainath and C.~Parada, ``Convolutional neural networks for small-footprint
  keyword spotting,'' in \emph{Interspeech}, 2015.

\bibitem{qian2016very}
Y.~Qian and P.~C. Woodland, ``Very deep convolutional neural networks for
  robust speech recognition,'' in \emph{2016 IEEE Spoken Language Technology
  Workshop (SLT)}.\hskip 1em plus 0.5em minus 0.4em\relax IEEE, 2016, pp.
  481--488.

\bibitem{arik2017convolutional}
S.~O. Arik \emph{et~al.}, ``Convolutional recurrent neural networks for
  small-footprint keyword spotting,'' \emph{arXiv:1703.05390}, 2017.

\bibitem{Tang2018}
J.~Tang, Y.~Song, L.~Dai, and I.~McLoughlin, ``Acoustic modeling with densely
  connected residual network for multichannel speech recognition,'' in
  \emph{Interspeech}, 2018.

\bibitem{choi2019temporal}
S.~Choi \emph{et~al.}, ``Temporal convolution for real-time keyword spotting on
  mobile devices,'' \emph{arXiv:1904.03814}, 2019.

\bibitem{kusupatiFastGRNN2018}
A.~Kusupati \emph{et~al.}, ``{FastGRNN}: A fast, accurate, stable and tiny
  kilobyte sized gated recurrent neural network,'' in \emph{NIPS}, 2018.

\bibitem{he2019streaming}
Y.~He \emph{et~al.}, ``Streaming end-to-end speech recognition for mobile
  devices,'' in \emph{ICASSP}, 2019.

\bibitem{coucke2019dilatedgating}
A.~Coucke \emph{et~al.}, ``Efficient keyword spotting using dilated
  convolutions and gating,'' in \emph{ICASSP}, 2019.

\bibitem{park2019}
D.~S. {Park} \emph{et~al.}, ``{SpecAugment}: A simple data augmentation method
  for automatic speech recognition,'' \emph{arXiv:1904.08779}, 2019.

\bibitem{devries2017specutout}
T.~DeVries and G.~W. Taylor, ``Improved regularization of convolutional neural
  networks with cutout,'' \emph{arXiv:1708.04552}, 2017.

\bibitem{novograd2019}
B.~{Ginsburg} \emph{et~al.}, ``Stochastic gradient methods with layer-wise
  adaptive moments for training of deep networks,'' \emph{arXiv:1905.11286},
  2019.

\bibitem{he2019bag}
T.~He \emph{et~al.}, ``Bag of tricks for image classification with
  convolutional neural networks,'' in \emph{CVPR}, 2019, pp. 558--567.

\bibitem{micikevicius2017mixed}
P.~Micikevicius \emph{et~al.}, ``Mixed precision training,''
  \emph{arXiv:1710.03740}, 2017.

\bibitem{nemo2019}
O.~Kuchaiev \emph{et~al.}, ``{NeMo}: a toolkit for building ai applications
  using neural modules,'' \emph{arXiv:1909.09577}, 2019.

\bibitem{huang2016}
G.~Huang, Z.~Liu, L.~van~der Maaten, and K.~Q. Weinberger, ``Densely connected
  convolutional networks,'' \emph{arXiv:1608.06993}, 2016.

\bibitem{lin2018edgespeechnets}
Z.~Q. Lin, A.~G. Chung, and A.~Wong, ``{EdgeSpeechNets}: Highly efficient deep
  neural networks for speech recognition on the edge,''
  \emph{arXiv:1810.08559}, 2018.

\bibitem{won2020harmonic}
M.~Won, S.~Chun, O.~Nieto, and X.~Serra, ``Data-driven harmonic filters for
  audio representation learning,'' in \emph{ICASSP}, 2020.

\bibitem{lin2020training}
J.~Lin, K.~Kilgour, D.~Roblek, and M.~Sharifi, ``Training keyword spotters with
  limited and synthesized speech data,'' \emph{arXiv:2002.01322}, 2020.

\bibitem{li2020feature}
B.~Li, F.~Wu, S.-N. Lim, S.~Belongie, and K.~Q. Weinberger, ``On feature
  normalization and data augmentation,'' \emph{arXiv:2002.11102}, 2020.

\bibitem{de2018neural}
D.~C. de~Andrade, S.~Leo, M.~L. D.~S. Viana, and C.~Bernkopf, ``A neural
  attention model for speech command recognition,'' \emph{arXiv:1808.08929},
  2018.

\bibitem{font2013freesound}
F.~Font, G.~Roma, and X.~Serra, ``Freesound technical demo,'' in
  \emph{Proceedings of the 21st ACM international conference on Multimedia},
  2013, pp. 411--412.

\end{thebibliography}

\end{document}